\documentclass[twocolumn,showpacs,superscriptaddress]{revtex4}%
\usepackage{amsfonts}
\usepackage{amsmath}
\usepackage{amssymb}
\usepackage{graphicx}%
\providecommand{\U}[1]{\protect\rule{.1in}{.1in}}

\begin{document}
\title{Phonon spectrum, thermodynamic properties, and pressure-temperature phase diagram of uranium dioxide}
\author{Bao-Tian Wang}
\thanks{E-mail: wbt11129@sxu.edu.cn}
\affiliation{Institute of Theoretical Physics and Department of
Physics, Shanxi University, Taiyuan 030006, People's Republic of
China}\affiliation{LCP, Institute of Applied Physics and Computational
Mathematics, Beijing 100088, People's Republic of China}\affiliation{Department of Physics and Astronomy, Division of Materials Theory, Uppsala University, Box 516, SE-75120 Uppsala, Sweden}
\author{Ping Zhang}
\thanks{E-mail: zhang\_ping@iapcm.ac.cn}
\affiliation{LCP, Institute of Applied Physics and Computational
Mathematics, Beijing 100088, People's Republic of China}
\author{Raquel Liz\'{a}rraga}
\affiliation{Instituto de Ciencias F\'{\i}sicas y Matem\'{a}ticas, Universidad Austral de Chile, Valdivia, Chile}
\author{Igor Di Marco}
\affiliation{Department of Physics and Astronomy, Division of
Materials Theory, Uppsala University, Box 516, SE-75120 Uppsala,
Sweden}
\author{Olle Eriksson}
\thanks{E-mail: olle.eriksson@physics.uu.se}
\affiliation{Department of Physics and Astronomy, Division of Materials Theory, Uppsala University, Box 516, SE-75120 Uppsala, Sweden}


\begin{abstract}
We present a study of the structural phase transition and the
mechanical and thermodynamic properties of UO$_{2}$ by means of the
local density approximation (LDA)$+U$ approach. A phase transition
pressure of 40 GPa is obtained from theory at 0 K, and agrees well
with the experimental value of 42 GPa. The pressure-induced
enhancements of elastic constants, elastic moduli, elastic wave
velocities, and Debye temperature of the ground-state fluorite phase
are predicted. The phonon spectra of both the ground state fluorite
structure and high pressure cotunnite structure calculated by the
supercell approach show that the cotunnite structure is dynamically
unstable under ambient pressure. Based on the imaginary mode along
the $\Gamma$$-$$X$ direction and soft phonon mode along the
$\Gamma$$-$$Z$ direction, a transition path from cotunnite to
fluorite has been identified. We calculate the lattice vibrational
energy in the quasiharmonic approximation using both
first-principles phonon density of state and the Debye model. The
calculated temperature dependence of lattice parameter, entropy, and
specific heat agrees well with experimental observations in the low
temperature domain. The difference of the Gibbs free energy between
the two phases of UO$_{2}$ has predicted a boundary in the
pressure-temperature phase diagram. The solid-liquid boundary is
approximated by an empirical equation using our calculated elastic
constants.

\end{abstract}
\maketitle

\section{INTRODUCTION}
Due to its critical importance in the nuclear fuel cycle and to the
complex electronic structure arising from a partially occupied
5\emph{f} orbital, uranium dioxide (UO$_{2}$) has been studied
extensively in experiments
\cite{Schoenes,Baer,Idiri,YuTobin,Tobin,An} and computational
simulations
\cite{Brooks,Dudarev3,Boettger2,Kudin,Prodan2,Yin,Petit,Zhang2010,Geng2011}.
The 5\emph{f} electrons in UO$_{2}$ play a pivotal role in
understanding its electronic, thermodynamic, and magnetic properties
\cite{Santini}. Using density functional theory (DFT) with a
conventional exchange-correlation potential, e.g., the local density
approximation (LDA) or generalized gradient approximation (GGA), an
incorrect ferromagnetic (FM) conducting ground state of UO$_{2}$ was
observed \cite{Boettger2} due to an error produced by
underestimating the strong on-site Coulomb repulsion of the
5\emph{f} electrons. Similar problems have been confirmed in
previous investigations of NpO$_{2}$ \cite{WangNpO2} and PuO$_{2}$
\cite{Zhang2010} within the pure LDA/GGA schemes. Fortunately, for
PuO$_{2}$ a theory based on completely localized 5\emph{f} states
reproduced well the crystal field splittings as well as the magnetic
susceptibility \cite{Colarieti-Tosti}. The $f$$\rightarrow$$f$
antiferromagnetic (AFM) Mott-Hubbard  insulator nature of UO$_{2}$
has been well reproduced in LDA/GGA+\emph{U} \cite{Dudarev3}, the
hybrid density functional HSE (Heyd, Scuseria, and Enzerhof)
\cite{Prodan2}, the self-interaction corrected local spin-density
(SIC-LSD) approximation \cite{Petit}, and LDA plus dynamical
mean-field theory (DMFT) \cite{Yin2011} calculations, which properly
describe the photoelectron spectroscopy experiments
\cite{Schoenes,Baer}.

At ambient conditions, UO$_{2}$ crystallizes in a cubic fluorite
structure (\emph{Fm$\bar{3}$m}, No. 225) with cations located in a
face-centered cubic (fcc) structure and anions occupying tetrahedral
sites. Similar to the high-pressure behavior of ThO$_{2}$ and
PuO$_{2}$ \cite{Dancausse}, a recent hydrostatic compression
experiment \cite{Idiri} has shown that UO$_{2}$ also transforms to
the orthorhombic structure of cotunnite-type (\emph{Pnma}, No. 62)
at room temperature, beyond 42 GPa. This kind of pressure-induced
phase transition (PT) for actinide dioxides is the same as for the
alkaline earth fluorides \cite{Dorfman} and has not been
sufficiently studied, although experiments \cite{Dancausse,Idiri}
and theoretical works \cite{Geng,WangThO2,Zhang2010} have paid great
attention to this issue. The data on the cotunnite phase are scarce
in the literature, especially for its thermodynamic properties and
vibrational characters. The temperature contributions to the PT have
not been included in previous studies. On the other hand, the
melting properties of UO$_{2}$ also have not been well investigated.
Only a few experiments have been conducted to describe the melting
of UO$_{2}$ near ambient pressure, because of the difficult
experimental conditions required to control and monitor the PT
\cite{Manara}.

In a previous systematic work \cite{Zhang2010}, the structural,
electronic, and mechanical properties of AFM UO$_{2}$ in its
ground-state fluorite phase were presented together with the
high-pressure cotunnite phase at their corresponding equilibrium
states, as given by LDA+\emph{U} with \emph{U}=4 eV. By means of the
third-order Birch-Murnaghan equation of state (EOS) \cite{Birch}
fitting, the lattice parameter $a_{0}$=5.449 \AA \ and bulk modulus
\emph{B}=220.0 GPa were found for \emph{Fm$\bar{3}$m} UO$_{2}$.
These values are in good agreement
with results of recent LDA+\emph{U} calculation \cite{Andersson} ($a_{0}%
$=5.448 \AA \ and \emph{B}=218 GPa), as well as experimental
investigations \cite{Yamashita,Idiri} ($a_{0}$=5.47 \AA \ and
\emph{B}=207 GPa). In the present work, we perform an extended study
of the structural, mechanical, and thermodynamic properties of
UO$_{2}$ in the pressure range from 0 to 250 GPa and in a
temperature interval from 0 to 4000 K. To this aim, we employ the
LDA+\emph{U} and GGA+\emph{U} schemes as implemented by Dudarev
\emph{et al}. \cite{Dudarev1,Dudarev2,Dudarev3}. The total energies
of the nonmagnetic (NM), AFM, and FM phases of the fluorite
structure have been calculated in a wide range of the effective
Hubbard \emph{U} parameter to check the validity of the ground-state
calculations. At 0 K, a $Fm\bar{3}m$$\rightarrow$$Pnma$ PT pressure
of 40 GPa is predicted. In addition, we have calculated the elastic
constants, elastic moduli, Poisson's ratio, elastic wave velocities,
and Debye temperature of AFM fluorite UO$_{2}$ in the pressure range
from 0 to 40 GPa. The structural transition path of the cotunnite
phase to the fluorite phase as well as the melting behavior, have
been studied based upon our calculated phonon dispersions, Gibbs
free energy, and elastic constants. Thermodynamic properties
including the Gibbs free energy, the temperature dependence of the
lattice parameter and the bulk modulus, entropy, and specific heat
have also been evaluated. The rest of this paper is arranged as
follows. In Sec. II the computational methods are described. In Sec.
III we present and discuss our results. In Sec. IV we summarize the
conclusions of this work.

\section{computational methods}

\subsection{Computational details}

First-principles DFT calculations are performed by means of the
Vienna \textit{ab initio} simulation package (VASP) \cite{Kresse3},
based on the frozen-core projected augmented wave (PAW) method of
Bl\"{o}chl \cite{PAW}. The exchange and correlation effects are
described with the LDA and GGA \cite{LDA,GGA}, and a cutoff energy
of 500 eV is used for the set of plane waves. The \emph{k}-point
meshes in the full wedge of the Brillouin zone (BZ) are sampled by
9$\times$9$\times$9 and 9$\times$15$\times$9 grids, respectively for
fluorite and cotunnite UO$_{2}$, according to the Monkhorst-Pack
(MP) \cite{Monk} scheme. All atoms are fully relaxed until the
Hellmann-Feynman (HF)
forces become less than 0.02 eV/\AA . The U 6s$^{2}$7s$^{2}$6p$^{6}%
$6d$^{2}$5f$^{2}$ and the O 2\emph{s}$^{2}$2\emph{p}$^{4}$ orbitals
are treated as valence electrons. Similar to our previous studies
\cite{WangNpO2,Zhang2010}, the strong on-site Coulomb repulsion
among the localized U 5\emph{f} electrons is described by using the
LDA/GGA+\emph{U} formulated by Dudarev \emph{et al.}
\cite{Dudarev1,Dudarev2,Dudarev3}, where the double counting
correction has already been included as in the fully localized limit
(FLL) \cite{liechtenstein95}. In this paper, we study several values
of the Hubbard parameter $U$, while we keep the Hund's exchange
parameter fixed to $J$=0.51 eV, following the results of Dudarev
\emph{et al.} \cite{Dudarev1,Dudarev2,Dudarev3}. One can notice that
only the difference between $U$ and $J$ is significant in our method
\cite{Dudarev2}, and we will henceforth refer to it as a single
parameter, named $U$ for sake of simplicity.

We calculate the ground-state properties of both phases of UO$_{2}$
by means of LDA/GGA+$U$ with and without the inclusion of spin-orbit
coupling (SOC). We find the AFM state to be lower in energy than the
FM state, which is in agreement with experimental observations and
with other calculations, as properly analyzed below. Then we
calculate elastic constants, phonon spectra and thermodynamics
properties at different pressures. These quantities are known to be
well described without including SOC for both UO$_{2}$ and PuO$_{2}$
\cite{Boettger1,Boettger2,ProdanJCP,Prodan1,Zhang2010,Sanati}. The
reason for this is that the 5\emph{f} states are chemically inert in
UO$_{2}$, due to their high localization \cite{Brooks}. The entire
chemical binding is provided by the \emph{spd} states of U and the
\emph{sp} states of O, and for these states SOC is less important.
Therefore, in most of our work on UO$_{2}$, the SOC is not included,
but we make a proper comparison to verify this approximation.

Additionally, in order to check the validity of our results, we
perform LDA+$U$ calculations with Elk \cite{elk}, a full-potential
augmented plane wave (FLAPW) method code. Here SOC is included for
magnetic calculations in a second-variational scheme, and the double
counting is chosen in the FLL. The muffin-tin (MT) radii (R$_{MT}$)
of U and O are set to 1.2 and 0.9 \AA,~ respectively. The parameter
R$_{MT}|\vec{G}+\vec{k}|_{\text{max}}$, which determines the number
of plane waves in the FLAPW method, is set to 9.5. A
$10\times10\times10$ grid is used to sample the BZ.

The Elk results are consistent with VASP, and the AFM configuration
is found to be the most energetically favorable state. In Elk, we
also calculate the total energy of the 3$\vec{k}$ magnetic
configuration, in which the star of the wave vector $\vec{k}$ of the
magnetic structure contains three members. The AFM configuration
(1$\vec{k}$) with magnetic moments aligned along the $z$ axis
(longitudinal) and within the $ab$ plane (transversal) are collinear
structures, whereas the 3$\vec{k}$ configurations (transversal and
longitudinal) are non-collinear. We compare the total energies and
find that the 1$\vec{k}$ configuration is the most stable one. The
3$\vec{k}$ longitudinal and transversal configurations are almost
degenerate, differing by only a few meV/U atom.

\subsection{Elastic properties, Debye temperature, and melting temperature}

To avoid the Pulay stress problem, the geometry optimization at each
volume is performed with VASP at fixed volume rather than constant
pressure. Elastic constants for cubic symmetry ($C_{11}$, $C_{12}$,
and $C_{44}$) and orthorhombic structure ($C_{11}$, $C_{12}$,
$C_{13}$, $C_{22}$, $C_{23}$, $C_{33}$, $C_{44}$, $C_{55}$, and
$C_{66}$) are calculated by applying stress tensors with various
small strains onto the equilibrium structures. The strain amplitude
$\delta$ is varied in steps of 0.006 from $\delta$=$-$0.036 to
0.036. A detailed description of the calculation scheme used here
can be found in Ref. \cite{Zhang2010}. After having obtained the
elastic constants, the polycrystalline bulk modulus \emph{B} and
shear modulus \emph{G} are calculated from the Voigt-Reuss-Hill
(VRH) approximations \cite{Hill}. The Young's modulus \emph{E} and
Poisson's ratio $\nu$ are calculated through $E=9BG/(3B+G)$ and
$\nu=(3B-2G)/[2(3B+G)]$. In the calculation of the Debye temperature
($\theta_{D}$), we use the relation
\begin{align}\label{debye_temp}
\theta_{D}=\frac{h}{k_{B}}\left(  \frac{3n}{4\pi\Omega}\right)  ^{1/3}%
\nu_{m},
\end{align}
where \emph{h} and $\emph{k}_{B}$ are Planck and Boltzmann
constants, respectively, \emph{n} is the number of atoms in the
molecule, $\Omega$ is molecular volume, and $\upsilon_{m}$ is the
average sound wave velocity. The average wave velocity in the
polycrystalline materials is approximately given by
\begin{align}
\upsilon_{m}=\left[  \frac{1}{3}\left(
\frac{2}{\upsilon_{t}^{3}}+\frac {1}{\upsilon_{l}^{3}}\right)
\right]  ^{-1/3},
\end{align}
where $\upsilon_{t}=\sqrt{G/\rho}$ ($\rho$ is the density) and
$\upsilon _{l}=\sqrt{(3B+4G)/3\rho}$ are the transverse and
longitudinal elastic wave velocity of the polycrystalline materials,
respectively. The melting temperature ($T_{m}$) in units of K for
cubic UO$_{2}$ is deduced from the elastic constant ($C_{11}$) by an
approximate empirical formula \cite{Fine}:
\begin{align}
T_{m}=553\text{ K}+\frac{5.91 \text{K}}{\text{GPa}}C_{11},
\end{align}
where the $C_{11}$ is in units of GPa and the standard error is about $\pm$300 K.

\subsection{Phonon and thermodynamic properties}

We use the supercell approach \cite{Parlinski} and the small
displacement method as implemented in the FROPHO code \cite{fropho}
to calculate the phonon curves in the BZ and the corresponding
phonon density of states (DOS) for both fluorite and cotunnite
phases of UO$_{2}$. In the interpolation of the force constants for
calculating the phonon dispersion, we sample the BZ of the
$Fm\bar{3}m$ 2$\times$2$\times$2 and $Pnma$ 2$\times$2$\times$2
supercells with respectively 3$\times$3$\times$3 and 3$\times
$5$\times$3 \emph{k} points. These meshes are set up by means of the
MP scheme. The forces induced by small displacements are calculated
within VASP.

Thermodynamic properties can be determined by phonon calculations
using the quasiharmonic approximation (QHA) \cite{Siegel,Zhang2010}
or the quasiharmonic Debye model \cite{GIBBS}. Within these two
models, the Gibbs free energy $G(T,P)$ is written as
\begin{equation}
G(T,P)=F(T,V)+PV.
\end{equation}
Here, $F(T,V)$ is the Helmholtz free energy at temperature \emph{T}
and volume \emph{V}, and can be expressed as
\begin{equation}
F(T,V)=E(V)+F_{vib}(T,V)+F_{el}(T,V),
\end{equation}
where $E(V)$ is the ground-state total energy, $F_{vib}(T,V)$ is the
vibrational energy of the lattice ions and $F_{el}(T,V)$ is the
thermal electronic contribution. Since we are treating a wide gap
insulator, we can avoid considering $F_{el}(T,V)$, as explained in
similar works \cite{Zhang2010}.

Under QHA, $F_{vib}(T,V)$ can be calculated by
\begin{equation}
F_{vib}(T,V)=k_{B}T\int_{0}^{\infty}g(\omega)\ln\left[  2
\sinh\left( \frac{\hslash\omega}{2k_{B}T}\right)  \right]  d\omega,
\end{equation}
where $\omega$ represents the phonon frequencies and $g(\omega)$ is
the phonon DOS. This formula strictly requires that the phonon DOS
is positive, and therefore it is not suitable for dynamically
unstable phases. In this case, the vibration energy for phases where
the phonon frequencies are imaginary can be estimated by the Debye
model
\begin{equation}
F_{vib}(T,V)=\frac{9}{8}k_{B}\theta_{D}+k_{B}T\left[3\ln\left
(1-e^{-\frac{\theta_{D}}{T}}\right) -D\left
(\frac{\theta_{D}}{T}\right)\right],
\end{equation}
where $\frac{9}{8}k_{B}\theta_{D}$ is the zero-point energy due to
lattice ion vibration at 0 K and $D(\theta_{D}/T)$ is the Debye
integral written as
$D(\theta_{D}/T)=3/(\theta_{D}/T)^{3}\int_{0}^{\theta_{D}/T}x^{3}/(e^{x}-1)dx$.
Note that $\theta_{D}$ here is not calculated by means of Eq.
(\ref{debye_temp}), but using a different prescription. For a more
detailed overview of the computational details, we redirect the
reader to Ref. \cite{GIBBS}.

\section{results}
\subsection{Phase transition at 0 K}

\begin{figure}[ptb]
\begin{center}
\includegraphics[width=1.0\linewidth]{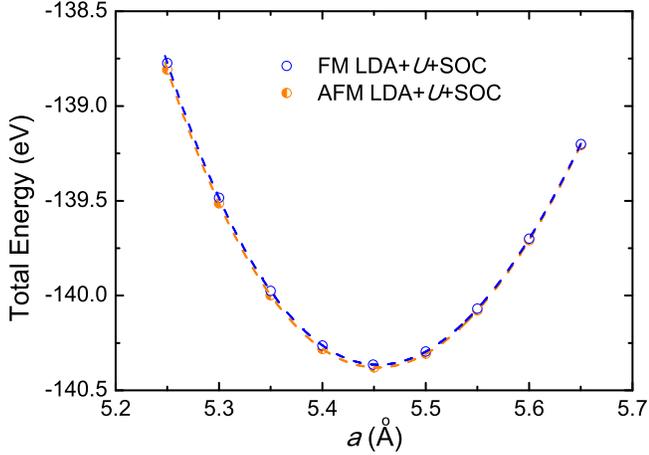}
\end{center}
\caption{(Color online) Total energy vs lattice constant $a$ for FM
and AFM UO$_{2}$ (12-atom cell) through LDA+\emph{U}+SOC with
\emph{U}=4 eV. The dashed lines are obtained from the EOS fitting.}%
\label{eos}%
\end{figure}

\begin{figure}[ptb]
\begin{center}
\includegraphics[width=1.0\linewidth]{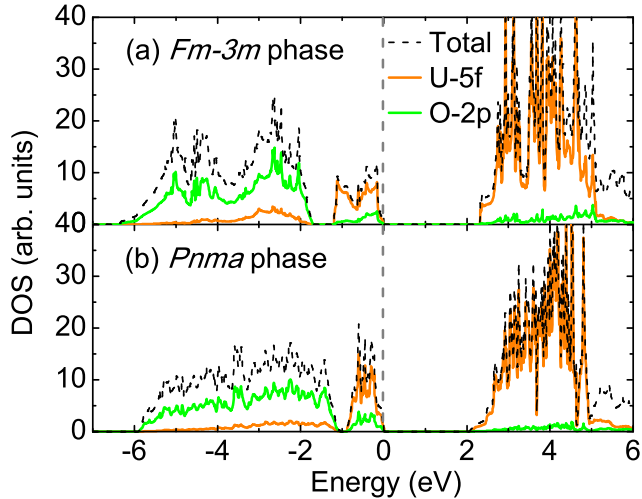}
\end{center}
\caption{(Color online) The total DOS and the projected DOSs of the
U 5$f$ and O 2$p$ orbitals for AFM UO$_{2}$ by LDA+\emph{U}+SOC with
\emph{U}=4 eV. In (a) and (b) we show the $Fm\bar{3}m$ and $Pnma$
phases respectively. The calculations are done at their
corresponding equilibrium volumes, optimized
within VASP.}%
\label{dos}%
\end{figure}

\begin{figure}[ptb]
\begin{center}
\includegraphics[width=1.0\linewidth]{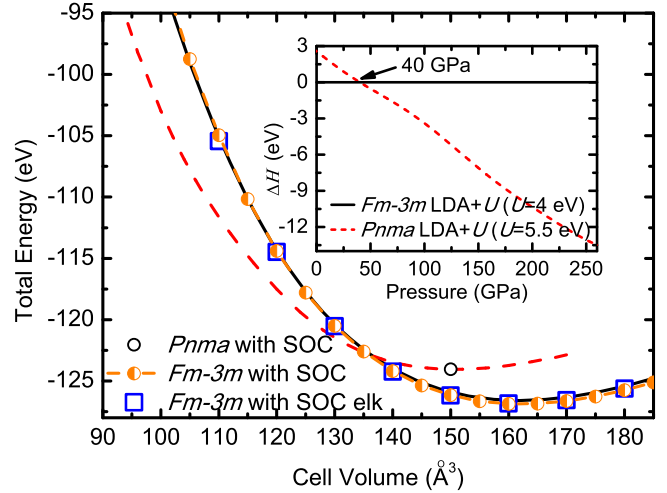}
\end{center}
\caption{(Color online) Total energy vs the cell volume for AFM
UO$_{2}$ in \emph{Fm$\bar{3}$m} and \emph{Pnma} phases. Results of
\emph{Fm$\bar{3}$m} are calculated by LDA+\emph{U} and
LDA+\emph{U}+SOC with \emph{U}=4 eV, while for \emph{Pnma} phase are
obtained by \emph{U}=5.5 eV. The 1$\vec{k}$ AFM equation of state of
\emph{Fm$\bar{3}$m} phase is also calculated by FLAPW method and SOC
and it is shown by hollow squares to compare with the VASP results.
Results with SOC have been moved into the same amplitude in the
energy level to better compare with the results without including
SOC. A PT at 40 GPa is predicted by the pressure dependence of the
enthalpy differences of the
$Pnma$ phase with respect to the $Fm\bar{3}m$ phase, as shown in the inset.}%
\label{energy}%
\end{figure}

In Fig. \ref{eos} we report the energy versus lattice constant
curves of the \emph{Fm$\bar{3}$m} phase in the AFM and FM
configurations, as obtained through VASP with LDA+\emph{U}+SOC and
\emph{U}=4 eV. As one can clearly observe, the AFM arrangement has
the lowest energy, and the energy difference with the FM arrangement
is 3.7 meV, which becomes 1.5 meV if GGA+\emph{U}+SOC is used. These
values are very consistent with recent DFT+\emph{U}+SOC results,
where an energy difference of about 6 meV is predicted \cite{Zhou},
with a slightly different $U$. Our results are also consistent with
the experimental N\'{e}el temperature of \emph{T}$_{N}$=30.8 K
\cite{Frazer}. By fitting our data for AFM configuration with the
EOS, we obtain an equilibrium lattice constant $a$ and a bulk
modulus $B$ of 5.453 \AA\ and 221 GPa, respectively. Instead, using
Elk with similar setup and $U$=4 eV, the optimized equilibrium
volume, lattice constant, and bulk modulus are 162.0 {\AA }$^{3}$,
5.440 {\AA }, and 230 GPa, respectively. These values are in good
agreement with our VASP calculations, and therefore support their
reliability.

Furthermore, we can evaluate the spin and orbital contribution to
the magnetic moment ($\mu_{s}$ and $\mu_{l}$). Our LDA+\emph{U}+SOC
calculations with VASP give values of $\mu_{s}$=1.30 $\mu_{B}$ and
$\mu_{l}$=-3.32 $\mu_{B}$ for the 1 \emph{k} AFM structure. These
are in reasonable agreement with previous DFT+\emph{U}+SOC values of
$\mu_{s}$=1.75 $\mu_{B}$ and $\mu_{l}$=-3.55 $\mu_{B}$ by the all
electron code WIEN2K \cite{Laskowski} and the experimental total
magnetic moment ($\mu_{\rm{total}}$) value of 1.74 $\mu_{B}$
\cite{FaberPRB}.

The total electronic DOS as well as the projected DOS for the U
5\emph{f} and O 2\emph{p} orbitals obtained by LDA+\emph{U}+SOC for
the \emph{Fm$\bar{3}$m} and the $Pnma$ phases are shown in Fig.
\ref{dos}. For the latter, the optimized structural lattice
parameters \emph{a}, \emph{b}, and \emph{c} are equal to 5.974,
3.605, and 6.965 {\AA } respectively, in the AFM configuration. The
energy band gaps ($E_{g}$) for the \emph{Fm$\bar{3}$m} and $Pnma$
phases are 2.3 and 2.0 eV, respectively. Our result for the
ground-state fluorite phase is in good agreement with the value of
$E_{g}$=2.4 eV, obtained in recent calculations with HSE+SOC
\cite{Wen}. However these numbers are still larger than the
experimental value of $E_{g}$=2.0 eV that was measured above the
N\'{e}el temperature \cite{Baer}.

\begin{table*}[ptb]
\caption{Lattice constants ($a$), elastic constants
(\emph{C$_{11}$}, \emph{C$_{12}$}, \emph{C$_{44}$}), bulk modulus
($B$), shear modulus ($G$), Young's modulus (\emph{E}), Poisson's
ratio ($\nu$), density ($\rho$), transverse ($\upsilon_{t}$),
longitudinal ($\upsilon_{l}$) and average ($\upsilon_{m}$) sound
velocities, and Debye temperature ($\theta_{D}$) for
\emph{Fm$\bar{3}$m} AFM UO$_{2}$ at different pressures calculated
through LDA+\emph{U} with \emph{U}=4 eV. For comparison,
experimental values and results from other calculations at 0 GPa are also listed.}%
\label{elastic}
\begin{ruledtabular}
\begin{tabular}{cccccccccccccccc}
Pressure&a&\emph{C$_{11}$}&\emph{C$_{12}$}&\emph{C$_{44}$}&\emph{B}&\emph{G}&\emph{E}&$\nu$&$\rho$&$\upsilon_{t}$&$\upsilon_{l}$&$\upsilon_{m}$&$\theta_{D}$\\
(GPa)&(\AA)&(GPa)&(GPa)&(GPa)&(GPa)&(GPa)&(GPa)&&(g/cm$^{3}$)&(m/s)&(m/s)&(m/s)&(K)\\
\hline
0&5.449&389.3&138.9&71.3&222.4&89.5&236.8&0.323&11.084&2841.8&5552.7&3183.4&398.1\\
5&5.408&414.8&154.5&94.3&241.2&107.3&280.4&0.306&11.343&3076.1&5821.1&3438.7&433.3\\
10&5.373&438.2&166.7&106.7&257.2&117.5&305.9&0.302&11.565&3187.5&5982.3&3561.2&451.7\\
15&5.340&459.2&181.6&118.9&274.1&126.5&328.9&0.300&11.776&3277.7&6131.8&3661.1&467.2\\
20&5.310&479.8&195.6&131.3&290.4&135.5&351.8&0.298&11.979&3363.3&6270.7&3755.8&482.0\\
25&5.282&500.3&208.3&143.6&305.6&144.6&374.6&0.296&12.175&3445.7&6398.0&3846.7&496.3\\
30&5.254&520.8&221.6&156.0&321.3&153.4&397.0&0.294&12.364&3522.3&6521.7&3931.4&509.8\\
35&5.229&540.0&233.7&167.8&335.8&161.8&418.2&0.292&12.546&3591.1&6630.0&4007.3&522.2\\
40&5.204&558.0&246.7&180.9&350.4&170.3&439.8&0.291&12.722&3659.1&6737.7&4082.5&534.5\\
Expt.&5.473$^{\emph{a}}$&389.3$^{\emph{b}}$&118.7$^{\emph{b}}$&59.7$^{\emph{b}}$&209.0$^{\emph{b}}$&83.0$^{\emph{b}}$&221.0$^{\emph{b}}$&0.324$^{\emph{b}}$&&&&&385$^{\emph{b}}$, 395$^{\emph{c}}$\\
LDA+\emph{U}$^{\emph{d}}$&5.448&380.9&140.4&63.2&220.6&82.0&218.9&0.335&&&&&399\\
\end{tabular}
$^{\emph{a}}$ Reference \cite{Idiri}, $^{\emph{b}}$ Reference
\cite{Fritz}, $^{\emph{c}}$ Reference \cite{Dolling}, $^{\emph{d}}$
Reference \cite{Sanati}.
\end{ruledtabular}
\end{table*}

Up to now, we have only presented results by DFT+\emph{U}+SOC.
However, our main focus in the present study is on the mechanical
properties, phonon spectrum, and thermodynamic properties. The
effect of the SOC on these quantities is rather small, as was
pointed out in Ref. \cite{Sanati}. Therefore, in the following, we
will present results obtained without SOC, and we will discuss the
associated errors, if relevant.

Using LDA+\emph{U} with \emph{U}=4 eV, we obtain $a$=5.449 \AA\ and
$B$=220 GPa for the \emph{Fm$\bar{3}$m} phase in AFM configuration
by EOS fitting. These values are identical to our previous results
\cite{Zhang2010}, and in good agreement with the corresponding
values by LDA+\emph{U}+SOC. The energy band gap and the spin
magnetic moment are calculated to be 1.9 eV and 1.98 $\mu_{B}$,
respectively. These values are in excellent agreement with both a
previous LDA+\emph{U} calculation \cite{Geng} (\emph{E}$_{g}$=1.45
eV and $\mu_{s}$=1.93 $\mu_{B}$) and experiments (\emph{E}$_{g}$=2.0
eV \cite{Baer}). Notice that here a comparing of the total magnetic
moment with experiments is not suitable due to the lack of the
relevant orbital contribution. For \emph{Pnma} UO$_{2}$ in AFM
phase, we obtain the optimized structural lattice parameters
\emph{a}, \emph{b}, and \emph{c} to be 5.974, 3.604, and 6.967 {\AA
}, respectively. The band gap is calculated to be 1.6 eV. Thus, the
band gap should not increase from the \emph{Fm$\bar{3}$m} phase to
the \emph{Pnma} phase either by LDA+\emph{U}+SOC or by LDA+\emph{U}.
This result is different from a previous LDA+\emph{U} calculation
\cite{Geng}, where an increase of the band gap was found at a cell
volume close to the transition pressure from 0.8 eV in the fluorite
phase to 2.4 eV in the cotunnite phase, by using different values of
Hubbard parameters.

In Fig. \ref{energy}, we show the total energy vs cell volume curves
of the \emph{Fm$\bar{3}$m} and \emph{Pnma} phases. These curves are
important for describing the PT under an externally applied
pressure. If one uses the same Hubbard parameter $U$ for both
phases, a PT is predicted at $\mathtt{\sim}$7 GPa, which is not
consistent with experimental data. It was previously argued
\cite{Geng} that a better description of the PT can be obtained by
using $U$= 5.5 eV, and in the present study we followed the
suggested prescription. From Fig. \ref{energy} it is clear that the
\emph{Fm$\bar{3}$m} phase is stable at ambient conditions and that a
transition to the \emph{Pnma} phase is expected under compression.
In the inset of Fig. \ref{energy} we show the relative enthalpies
$H$ of the \emph{Pnma} phase with respect to the \emph{Fm$\bar{3}$m}
phase as a function of the pressure. Considering that at 0 K the
Gibbs free energy is equal to the enthalpy, we can then identify the
PT pressure as 40 GPa, as indicated by the cross point. This is
consistent with the previous LDA+$U$ results of about 38 GPa
\cite{Geng}, and also with the experimentally observed value of 42
GPa \cite{Idiri}. Finally in Fig. \ref{energy} we also show results
obtained with LDA+\emph{U}+SOC with VASP and with Elk. The good
agreement that one can observe between the three sets of simulations
indicates that the effects associated with the SOC can be neglected
when calculating the elastic and structure properties of UO$_{2}$.
\subsection{Elasticity of fluorite UO$_{2}$}

The elastic constants can measure the resistance and mechanical
properties of a crystal under external stress or pressure, thus
describing the stability of crystals against elastic deformation. We
present in Table I the lattice constant, elastic constants, bulk
modulus, shear modulus, Young modulus, Poisson's ratio, density,
elastic wave velocities, and Debye temperature for
\emph{Fm$\bar{3}$m} AFM UO$_{2}$ at different pressures. All these
values are obtained through LDA+\emph{U} VASP calculations with
\emph{U}=4 eV. We also calculate the elastic constants at 0 GPa by
including SOC using VASP, obtaining $C_{11}$=395.9 GPa,
$C_{12}$=134.0 GPa, and $C_{44}$=89.5 GPa. These values are in close
agreement with both our LDA+\emph{U} results and experiments, as
shown in Table I, and illustrate that the inclusion of SOC is not
crucial for the elastic properties of UO$_{2}$. Elastic constants at
0 GPa have been widely studied by experiments \cite{Fritz} or
through first-principles calculations
\cite{Dudarev_elastic,Devey,Sanati}. Our calculated results at zero
pressure are consistent with these values, and in particular with
the recent LDA+\emph{U} work of Sanati \emph{et al.} \cite{Sanati}.
There, the author also show that the SOC introduces only marginal
changes in the elastic properties of UO$_{2}$, thus supporting our
chosen methodology for this study. In the entire pressure range
considered in our study, $C_{11}$ is prominently larger than
$C_{12}$, indicating that the bonding strength along the
[100]/[010]/[001] directions is clearly stronger than that of the
bonding along the [011]/[101]/[110] directions. In fact, there are
eight U-O covalent bonds per formula unit for fluorite UO$_{2}$. The
angle of all eight bonds with respect to the [100]/[010]/[001]
directions is 45$^{\circ}$. However, only four bonds make an angle
of 45$^{\circ}$ with the [011]/[101]/[110] directions. Four other
bonds are vertical to the strain directions of [011]/[101]/[110],
and they have no contributions to the elastic strength. Therefore,
it is intuitive that $C_{11}>C_{12}$ for cubic UO$_{2}$. This kind
of analysis of the chemical bonding has been previously used to
explain the different theoretical tensile strengths in the three
typical crystalline orientations of PuO$_{2}$ \cite{Zhang2010}.
Finally, for the Debye temperature, our calculated result of 398.1 K
is in excellent agreement with experimental data
\cite{Fritz,Dolling}.

As indicated in Table I, pressure-induced enhancements of elastic
constants, elastic moduli, elastic wave velocities, and Debye
temperatures are evident with the exception of the Poisson's ratio.
These quantities all increase linearly with pressure. While $C_{12}$
and $C_{44}$ have the same increase rate of $\mathtt{\sim}$2.7,
$C_{11}$ has a larger one of $\mathtt{\sim}$4.2. This can also be
understood from the previous bonding analysis. The rates with which
\emph{B}, \emph{G}, and \emph{E} increase, are 3.2, 2.0, and 5.1,
respectively. Considering that $B_{V}=B_{R}=(C_{11}+2C_{12})/3$,
$G_{V}=(C_{11}-C_{12}+3C_{44})/5$, and
$G_{R}=5(C_{11}-C_{12})C_{44}/[4C_{44}+3(C_{11}-C_{12})]$ for cubic
symmetry, we can understand why the increase rate of \emph{G} is
only about 60\% of the increase rate of \emph{B}. For transverse
($\upsilon_{t}$) and longitudinal ($\upsilon_{l}$) sound velocities,
increase rates of 20.4 and 29.6 m s$^{-1}$GPa$^{-1}$ are
respectively obtained. The larger increase rate of the transverse
sound velocity upon compression is due to the larger enhancement of
the bulk modulus \emph{B} with respect to the shear modulus
\emph{G}. The linear increase of the Debye temperature under
pressure is also evident from this analysis, and can supply useful
informations in practical applications and/or theoretical
investigations of UO$_{2}$.

\subsection{Phonon dispersion}

\begin{figure}[ptb]
\begin{center}
\includegraphics[width=1.0\linewidth]{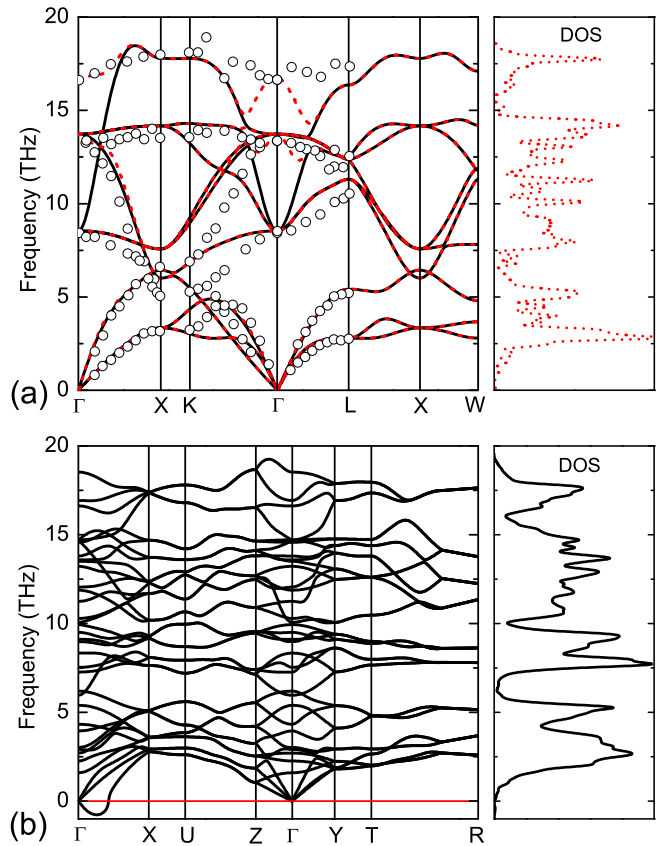}
\end{center}
\caption{(Color online) Phonon dispersion curves (left panel) and
corresponding phonon DOS (right panel) for UO$_{2}$ in (a)
$Fm\bar{3}m$ phase and (b) $Pnma$ phase. All results are calculated
through LDA+\emph{U} with \emph{U}=4 eV. In (a), solid and dashed
lines refer respectively to calculations without and with
polarization effects. The hollow circles present the experimental data from Ref. \cite{Dolling}.}%
\label{phonon}%
\end{figure}

The calculated phonon dispersion curves as well as the corresponding
phonon DOS are displayed in Fig. \ref{phonon} for $Fm\bar{3}m$ and
$Pnma$ UO$_{2}$ in the AFM configuration. To our knowledge, no
experimental or theoretical results on phonons have been published
for the high-pressure phase of actinide dioxides. For UO$_{2}$ in
the $Fm\bar{3}m$ phase, i.e., at ambient pressure, several
experimental techniques have been used to evaluate its vibrational
properties, namely inelastic neutron scattering \cite{Dolling,Pang}
and infrared and Raman spectroscopy
\cite{Schoenes,LivnehPRB,LivnehJPCM}. Also, from the theoretical
side, this system has been widely investigated, e.g., through
LDA+DMFT \cite{Yin}, MD \cite{Goel}, GGA \cite{Yun}, and
LDA/GGA+$U$+SOC \cite{Sanati}. In Fig. \ref{phonon}(a) we show the
phonon dispersion of the $Fm\bar{3}m$ phase along
$\Gamma$-$X$-$K$-$\Gamma$-$L$-$X$-$W$ directions. The segments
$\Gamma$-$X$, $\Gamma$-$K$, and $\Gamma$-$L$, are respectively along
the [001], [110], and [111] directions. Here we should note that
neglecting the SOC in plain LDA or GGA leads to underestimating the
optical modes, since the modes at high frequencies are shifted to
lower frequencies. However, in LDA+$U$ (for large enough $U$) this
problem disappears, and it has been proved that SOC does not
introduce any significant correction \cite{Sanati}. From Fig.
\ref{phonon}(a), one can find that including polarization effects is
necessary to correctly account for the longitudinal optical
(LO)-transverse optical (TO) splitting near the $\Gamma$ point in
BZ. Here, the Born effective charges ($Z_{\mathrm{{U}}}^{\ast}$=5.54
and $Z_{\mathrm{{O}}}^{\ast}$=$-$2.77) of U and O ions for
$Fm\bar{3}m$ AFM UO$_{2}$ are also calculated. Our phonon
dispersions are overall in good agreement with the inelastic neutron
scattering experiment \cite{Dolling,Pang} and previous calculations
\cite{Yin,Goel,Yun,Sanati}.

\begin{figure}[ptb]
\begin{center}
\includegraphics[width=0.8\linewidth]{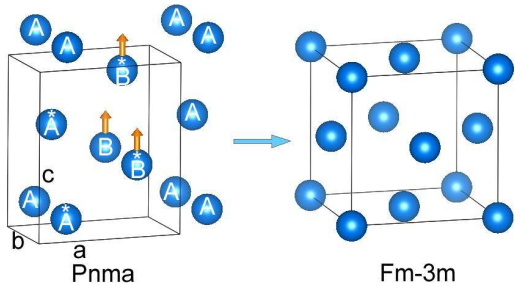}
\end{center}
\caption{(Color online) Schematic illustrations of the structural transition
from $Pnma$ phase to $Fm\bar{3}m$ structure. For clarity, only uranium atoms are presented and atoms within the $Pnma$ unit cell are labeled by star symbols.}%
\label{path}%
\end{figure}

In Fig. \ref{phonon}(b) we show the phonon dispersion of the $Pnma$
phase along $\Gamma$-$X$-$U$-$Z$-$\Gamma$-$Y$-$T$-$R$ directions.
The high-symmetry points here correspond to $\Gamma$ (0, 0, 0), $X$
(0, $\frac{1}{2}$, 0), $U$ (0, $\frac{1}{2}$, $\frac{1}{2}$), $Z$
(0, 0, $\frac{1}{2}$), $Y$ (-$\frac{1}{2}$, 0, 0), $T$
(-$\frac{1}{2}$, 0, $\frac{1}{2}$), and $R$ (-$\frac{1}{2}$,
$\frac{1}{2}$, $\frac{1}{2}$). Although in our previous work on the
elastic constants \cite{Zhang2010} we have predicted the $Pnma$
phase of UO$_{2}$ to be mechanically stable in its equilibrium
state, Fig. \ref{phonon}(b) clearly shows that the transverse
acoustic (TA) mode close to the $\Gamma$ point becomes imaginary
along the $\Gamma$-$X$ (i.e., the [010]) direction. This means that
the high-pressure phase of UO$_{2}$ is dynamically unstable at
ambient pressure. In addition, we can find a clear soft phonon mode
along the $\Gamma$-$Z$ (i.e., the [001]) direction. Thus, U atoms in
the $Pnma$ structure are easy to move along the [010] and [001]
directions. Based on these observations, we show in Fig. \ref{path}
a suggested path for the $Pnma$$\mathtt{\rightarrow}$$Fm\bar{3}m$
transition. The $Pnma$ phase can be viewed as an AB periodically
layered structure along the [100] direction. During the transition,
at the beginning the adjacent (100) planes slip relatively along the
[001] direction to create a face-centered orthorhombic structure (as
indicated by the arrows in Fig. \ref{path}). Then, the cell expands
along the [010] direction and shrinks in the vertical directions to
form the fcc fluorite structure.

\begin{figure}[ptb]
\begin{center}
\includegraphics[width=1.0\linewidth]{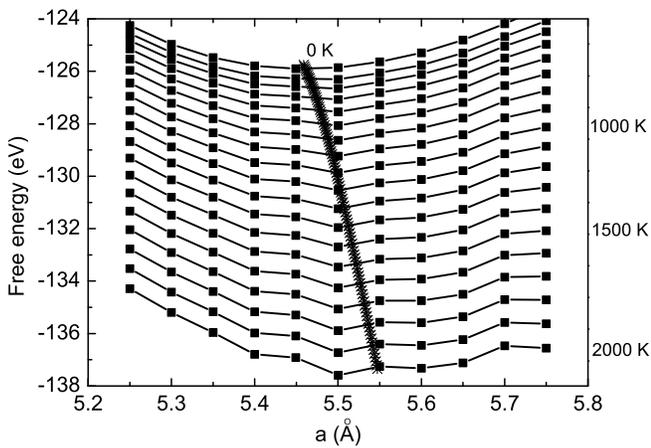}
\end{center}
\caption{Dependence of the free energy $F(T,V)$ on the lattice
parameter \emph{a} at various temperatures for AFM UO$_{2}$
calculated by LDA+\emph{U} with \emph{U}=4 eV.}%
\label{freeE}%
\end{figure}

\begin{figure}[ptb]
\begin{center}
\includegraphics[width=1.0\linewidth]{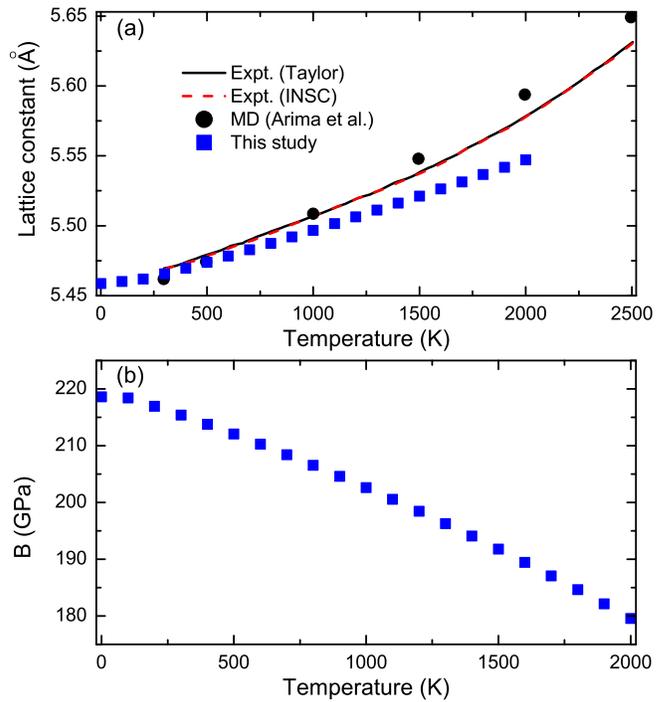}
\end{center}
\caption{(Color online) Temperature dependence of (a) lattice
parameter $a(T)$ and (b) bulk modulus \emph{B(T)} of UO$_{2}$.
Experimental results
from Refs. \cite{Taylor} and \cite{INSC} as well as the MD results from \cite{Arima} are also shown in panel (a).}%
\label{latticeT}%
\end{figure}

\subsection{Thermodynamic properties and $P-T$ phase diagram}

\begin{figure*}[ptb]
\begin{center}
\includegraphics[width=0.8\linewidth]{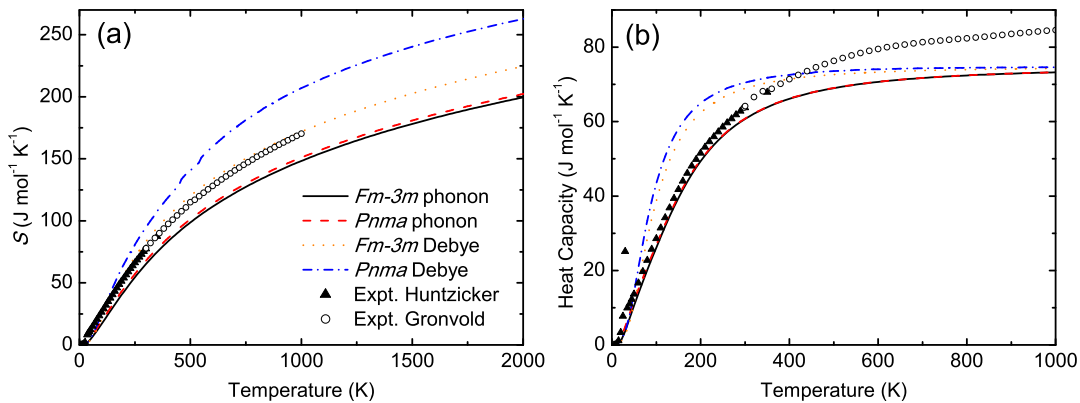}
\end{center}
\caption{(Color online) Temperature dependences of (a) entropy and (b) heat capacity at constant volume, $C_{V}$, for UO$_{2}$ in the $Fm\bar{3}m$ and $Pnma$ phases at 0 GPa. Results of the QHA and of the Debye model are presented together with experimental values from Refs. \cite{Gronvold} and \cite{Huntzicker}.}%
\label{Cv}%
\end{figure*}

\begin{figure*}[ptb]
\begin{center}
\includegraphics[width=0.8\linewidth]{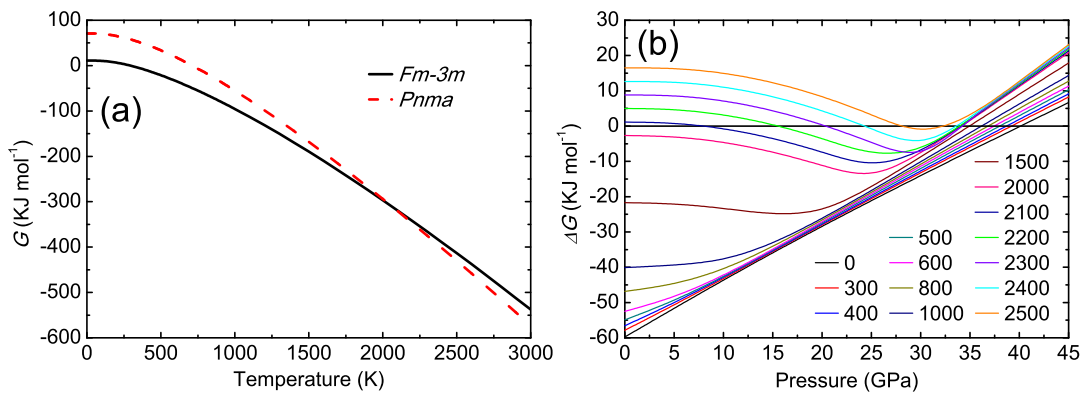}
\end{center}
\caption{(Color online) (a) Temperature dependence of the Gibbs free
energy for UO$_{2}$ in the $Fm\bar{3}m$ and $Pnma$ phases at 0 GPa.
(b) Difference of the Gibbs free energies of the $Fm\bar{3}m$ and
$Pnma$ phases of UO$_{2}$ as a function of pressure for
several temperatures (in K). A positive (negative) value indicates that the $Pnma$ ($Fm\bar{3}m$) phase is more stable.}%
\label{Gibbs}%
\end{figure*}

The calculated free energy curves $F(T,V)$ of UO$_{2}$ for
temperatures ranging from 0 up to 2000 K are shown in Fig.
\ref{freeE}. Note that in the calculation of $F(T,V)$, the
ground-state total energy and phonon free energy should be evaluated
by constructing several 2$\times $2$\times$2 fcc supercells. This
procedure is computationally very expensive. In Fig. \ref{freeE},
the equilibrium lattice parameters at different temperature \emph{T}
are also presented. Figure \ref{latticeT} shows the temperature
dependence of the lattice parameter and the bulk modulus. The
equilibrium volume $V(T)$ and the bulk modulus $B(T)$ are obtained
by EOS fitting. Experimental results from Refs. \cite{Taylor} and
\cite{INSC} as well as the MD results from \cite{Arima} are also
plotted. We observe a good agreement of the calculated lattice
parameters with respect to the experiments in the low-temperature
domain. However, our values are somewhat lower than the experimental
ones for temperatures higher than 800 K. The differences may come
from the thermal electronic contribution and/or anharmonic effects.
Similar to PuO$_{2}$ \cite{Zhang2010}, the bulk modulus $B(T)$
decreases when the temperature is increased. For UO$_{2}$ the
amplitude of such a change between 0 and 1500 K is
$\mathtt{\sim}$26.8 GPa, which is larger than that of PuO$_{2}$ by
about 6.2 GPa. This means that UO$_{2}$ will be softened quicker
upon increasing temperature in comparison with PuO$_{2}$.

Using the QHA and the Debye model, we have calculated the Gibbs free
energy ($G$), entropy ($S$), and specific heat at constant volume
($C_{V}$) for $Fm\bar{3}m$ and $Pnma$ phases of UO$_{2}$. Notice
that since the specific heat at constant pressure ($C_{P}$) has
similar trends as $C_{V}$ \cite{Zhang2010}, in the present work will
refer only to the latter. In Fig. \ref{Cv} the dependence of the
$C_{V}$ and $S$ on the temperature $T$ is showed, together with the
experimental results from Refs. \cite{Gronvold} and
\cite{Huntzicker}. Under the QHA, the curves of the entropy $S$ for
the $Fm\bar{3}m$ and $Pnma$ phases are almost identical to each
other. The $S$ of fluorite UO$_{2}$ is underestimated in a wide
range of temperatures with respect to the experiments, in agreement
with recent calculations \cite{Sanati}.


However, as clearly indicated in Fig. \ref{Cv}(a), the Debye model
can give proper results for $Fm\bar{3}m$ UO$_{2}$. Using the Debye
model, the $S$ curves for the $Fm\bar{3}m$ and $Pnma$ phases will
separate when increasing temperature. The difference between the QHA
and the Debye model is due to the fact that the Debye model includes
some anharmonic contributions in the calculation of $S$ and $C_{V}$,
while the QHA does not. Although the Debye model is less accurate,
it can supply a qualitative picture or even a quantitative
description of the thermodynamic properties. As shown in Fig.
\ref{Cv}(b), the $C_{V}$ of $Fm\bar{3}m$ UO$_{2}$ under the QHA
agrees well with experiments up to room temperature and becomes
close to a constant in the Dulong-Petit limit \cite{Kittel}. Similar
trends have been recently observed for the $C_{V}$ of the
$Fm\bar{3}m$ phase by Sanati \emph{et al.} \cite{Sanati}. Our
results point to that the $C_{V}$ curves for the $Fm\bar{3}m$ and
$Pnma$ phases are almost identical to each other. However, in the
Debye model, a slower increase of the $C_{V}$ when increasing the
temperature is observed for the $Fm\bar{3}m$ phase with respect to
the $Pnma$ phase. The Debye model gives $\theta_{D}$=390.6 and 352.8
K for the $Fm\bar{3}m$ and $Pnma$ phases respectively, and these
values are in good agreement with the values of 398.1 and 343.7 K
computed from the elastic constants \cite{Zhang2010}.

As shown in Fig. \ref{Gibbs}(a), the crossing between the Gibbs free
energy of the $Fm\bar{3}m$ and $Pnma$ phases, as obtained through
the Debye model, clearly gives a
$Fm\bar{3}m$$\mathtt{\rightarrow}$$Pnma$ PT temperature of 2069 K.
This implies a significant temperature contribution for the
$Fm\bar{3}m$$\mathtt{\rightarrow}$$Pnma$ PT, which is hence not only
pressure driven. To predict the phase boundary of this PT, we
calculate the Gibbs free energy of the $Fm\bar{3}m$ and $Pnma$
crystal structures in a temperature range from 0 to 3000 K, and the
effect of the pressure is studied in the range between 0 and 45 GPa.
The difference of the Gibbs energy ($\Delta$$G$) between the
fluorite and cotunnite structures of UO$_{2}$ as a function of
pressure for several temperatures is reported in Fig.
\ref{Gibbs}(b). At 0 K, the $Fm\bar{3}m$$\mathtt{\rightarrow}$$Pnma$
PT pressure is predicted to be 40 GPa, corresponding to our
aforementioned result. Along with increasing temperature in the
range from 0 to 2069 K, the pressure of the
$Fm\bar{3}m$$\mathtt{\rightarrow}$$Pnma$ transition decreases
slightly. At higher temperatures, the $Fm\bar{3}m$ phase is only
stable in middle pressure range.

\begin{figure}[ptb]
\begin{center}
\includegraphics[width=1.0\linewidth]{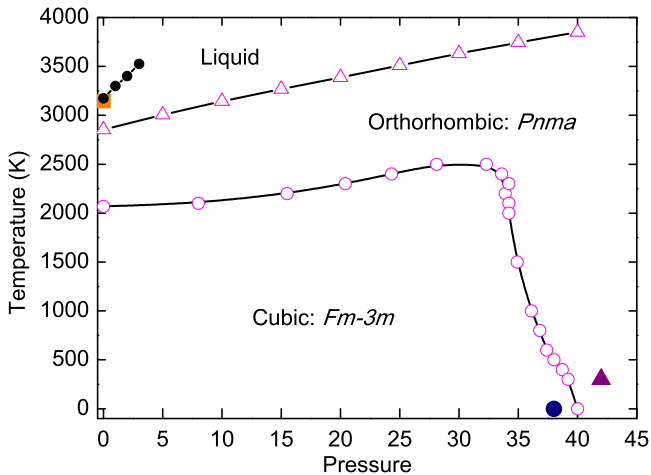}
\end{center}
\caption{(Color online) Calculated \emph{P-T} phase diagram for
uranium dioxide (indicated by line plus hollow symbols)
compared with selected experimental points [square for melting point (Res. \cite{Manara}) and triangle for PT pressure (Ref. \cite{Idiri})] and other calculations [line plus small solid circles for MD (Ref. \cite{Arima}) and the large circle for LDA+\emph{U} (Ref. \cite{Geng})].}%
\label{phasediagram}%
\end{figure}

Once the free energies of the two experimentally observed structures
are determined, the phase boundary can be obtained by equating the
Gibbs free energies at a given pressure and temperature. Besides,
the solid-liquid boundary can be featured by the melting temperature
$T_{m}$. Based on these results, we can plot in Fig.
\ref{phasediagram} the phase diagram of UO$_{2}$. Other theoretical
\cite{Arima,Geng} and experimental \cite{Manara,Idiri} values are
also presented for comparison. For the phase boundary between
$Fm\bar{3}m$ and $Pnma$, only one point at ambient condition was
reported in experiment. Our predicted results call for further
experimental and theoretical works, to investigate the accuracy of
the theory. For the solid-liquid boundary, the experimentally
determined melting value at zero pressure is 3147 K \cite{Manara}.
We note that experiment and recent MD calculation have reported the
relationship between melting point and pressure to be $T_{m,P}=$3147
K$+$$\frac{92.9\text{ K}}{\text{GPa}}$$P$ and $T_{m,P}=$3178
K$+$$\frac{115\text{ K}}{\text{GPa}}$$P$, respectively, where $P$ is
pressure in unit of GPa. Using our calculated data in Fig.
\ref{phasediagram}, we obtain $T_{m,P}=$2882 K$+$$\frac{24.8\text{
K}}{\text{GPa}}$$P$. The melting point at zero pressure is
underestimated by about 265 K, which is the same as previous LDA+$U$
calculations \cite{Sanati}. The increasing rate of $T_{m,P}$ is
largely underestimated compared to experiment. Although we cannot
claim that our calculations are more reliable than these
experiments, we note that the latter were performed only in a narrow
range of pressure, between 0.01 and 0.25 GPa.

\section{CONCLUSION}

In summary, we have carried out a first-principles DFT+\emph{U}
exploration of the ground-state properties as well as the
high-temperature/pressure behavior of UO$_{2}$ within VASP. For a
few selected cases these calculations have been compared with
LDA+$U$+SOC simulations within VASP and Elk. We find that all types
of calculations resulted in equilibrium volumes and elastic
constants which are in good agreement with each other. This shows
that the inclusion of SOC does not significantly influence the
structural properties of UO$_{2}$. By choosing the Hubbard \emph{U}
parameter around 4 eV within the LDA+\emph{U} approach, the
equilibrium state features for both the ambient \emph{Fm$\bar{3}$m}
and the high-pressure \emph{Pnma} phases of UO$_{2}$ are shown to
agree well with experiments. However, the
\emph{Fm$\bar{3}$m}$\mathtt{\rightarrow}$\emph{Pnma} transition is
predicted to occur at only 7 GPa, and only by increasing $U$ to 5.5
eV for the $Pnma$ phase can one find a value of 40 GPa, which is in
good agreement with experiment. This finding is also in good
agreement with a recent theoretical study \cite{Geng}. At ambient
pressure, a transition temperature of 2069 K between the two solid
structures is firstly obtained by the Debye model. The mechanical
properties and the Debye temperature of the fluorite phase have been
observed to increase linearly with the pressure, by calculating the
elastic constants. As a result, the melting temperature $T_{m}$ also
increases linearly with the pressure. Phonon dispersion results of
the \emph{Fm$\bar{3}$m} phase are in good agreement with available
experimental values. The LO-TO splitting at the $\Gamma$ point is
successfully reproduced by including the polarization effects. For
the cotunnite phase, the imaginary mode along the $\Gamma$-$X$
direction and soft phonon mode along the $\Gamma$-$Z$ direction have
been found at the equilibrium volume. The cotunnite to fluorite
transition can be reached by firstly slipping the adjacent (100)
planes relatively along the [001] direction to create a
face-centered orthorhombic structure and secondly expanding the cell
along the [010] direction and shrinking in the vertical directions
to form the fcc fluorite structure. Using the QHA and the Debye
model, we have calculated the Gibbs free energy, temperature
dependences of lattice parameter and bulk modulus, entropy, specific
heat, and $P-T$ phase diagram of UO$_{2}$. Given the importance of
this material as nuclear fuel we expect these results to be useful
for further theoretical and experimental investigations.

\begin{acknowledgments}
This work was supported by NSFC under Grant No. 11104170, No.
51071032, and No. 11074155, and by the Foundations for Development
of Science and Technology of China Academy of Engineering Physics
under Grant No. 2009B0301037. O.E. acknowledges support from the
Swedish research council, the KAW foundation, ESSENCE, STANDUPP, and
the ERC (Project No. 247062-ASD).
\end{acknowledgments}

\end{document}